\documentclass[aps,prl,twocolumn,floatfix,amsmath,superscriptaddress]{revtex4-1}

\usepackage{bm}
\usepackage{graphicx}
\usepackage{latexsym}
\usepackage{amssymb}
\usepackage{changes}
\usepackage{soul}
\usepackage{lineno}
\usepackage{xcolor}
   \usepackage{verbatim}           
 
\usepackage[version=4]{mhchem}

\usepackage{siunitx}
\DeclareSIUnit\torr{Torr}
\usepackage[
    colorlinks = true,         
    linkcolor  = blue,         
    citecolor  = blue,         
    urlcolor   = blue          
]{hyperref}

\usepackage{color}

\begin{document}
\title{Reconstructing Molecular-Ion Quantum States with Two-Color Strong-Field Scattering}

\author{Dina S. Eissa}
\email{deissa@iu.edu}
\affiliation{Department of Physics, The Ohio State University, Columbus, OH, USA}
\affiliation{Department of Physics, Indiana University, Bloomington, IN, USA}

\author{Felipe Morales}
\affiliation{Max Born Institute, Berlin, Germany} 
 
\author{Abraham Camacho Garibay}
\affiliation{Department of Physics, The Ohio State University, Columbus, OH, USA}
\affiliation{Instituto Politécnico Nacional, Mexico City, Mexico}

\author{Vyacheslav Leshchenko}
\affiliation{Department of Physics, The Ohio State University, Columbus, OH, USA}
\affiliation{Linac Coherent Light Source, SLAC National Accelerator Laboratory, Menlo Park, CA 94025, USA}

\author{Cosmin I. Blaga}
\affiliation{J. R. Macdonald Laboratory, Department of Physics, Kansas State University, Manhattan, Kansas, USA}

\author{Misha Ivanov}
\affiliation{Max Born Institute, Berlin, Germany} 

\author{Pierre Agostini}
\affiliation{Department of Physics, The Ohio State University, Columbus, OH, USA}

\author{Maria Richter}
\affiliation{Max Born Institute, Berlin, Germany} 

\author{Louis F. DiMauro}
\email{dimauro.6@osu.edu}
\affiliation{Department of Physics, The Ohio State University, Columbus, OH, USA}

\pacs{42.50.Hz, 32.80.-t, 32.80.Rm}

\begin{abstract} \noindent
Strong-field ionization of molecules can prepare coherent superpositions of ionic states whose amplitudes and relative phases govern the ensuing ultrafast dynamics, but accessing this information experimentally remains a challenge. Here we introduce two-color strong-field scattering, in which a strong mid-infrared field ionizes the molecule and launches an electron, while a weak second harmonic field resonantly couples vibronic states of the parent ion during the electron's continuum excursion, thereby introducing structural changes on the sub-laser-cycle timescale. Scanning the relative phase between the two colors controls the ionic evolution and maps it onto phase-dependent sub-\AA{} bond-length changes retrieved from the rescattering signal. Using the example of N$_2$, we apply the two-color setup to probe the coherence between different ionization channels and to reconstruct the initial ionic superposition. Our results establish two-color strong-field scattering as a powerful structural approach to characterize the complex amplitudes of ion quantum wavepackets. 
\end{abstract}

\maketitle

Strong-field ionization (SFI) can coherently populate multiple electronic states of the molecular ion, preparing an ionic wave packet whose populations and coherences set the initial conditions for the subsequent electronic and nuclear dynamics~\cite{Pabst2016PRA}. Characterizing these wave packets is important for understanding a broad range of processes, including charge migration in molecular ions~\cite{Kraus2015Science}, coupled electron-nuclear motion and electronic decoherence~\cite{Vacher2017PRL}, laser-driven nonadiabatic population transfer into dissociative ionic states~\cite{Zhao2017PRA}, and ultrafast mirrorless nitrogen-ion lasing in air~\cite{Yao2011,Richter2020Optica}. Experimentally, however, the initial ionic-state populations and relative phases are difficult to recover because measured observables reflect both the ionic state prepared during ionization and its subsequent field-driven evolution. Here, we combine a phase-sensitive structural observable with time-dependent simulations of the laser-driven ionic wave-packet dynamics to constrain the populations and relative phase of the initial ionic superposition prepared by SFI.

Previous attosecond and strong-field measurements have demonstrated access to the amplitudes and phases associated with multichannel ionization. High-harmonic generation (HHG) spectroscopy revealed multiorbital contributions to strong-field ionization in \ce{CO2} and \ce{N2} and enabled imaging of the  resulting electron--hole wave-packet dynamics~ \cite{Smirnova2009Nature,McFarland2008Science,Haessler2010NatPhys,Mairesse2010PRL}. Multidimensional HHG spectroscopy subsequently reconstructed the relative amplitudes and phases of \ce{CO2} ionization channels using phase-controlled  two-color fields~\cite{Bruner2016FD}.  More broadly, phase-controlled multicolor fields have become powerful tools for  steering and clocking sub-cycle electron dynamics in HHG and strong-field photoelectron spectroscopy~\cite{Brugnera2011PRL_TwoColorTrajectories,Niikura2010PRL_TwoColorOrbitalSymmetry,Skruszewicz2015,Chaloupka2016PRL,Porat2018NatComm,Piper2025PRL_QTS,Jin2024UltrafastScience}, while related interferometric approaches have enabled retrieval of continuum-channel and bound-state phase information~
\cite{Zipp2014,Liu2015PRL,Peschel2022NatCommun}.

Strong-field rescattering provides a natural framework for probing ultrafast molecular-ion dynamics: an electron is tunnel-ionized by an intense laser field and driven back to its parent ion within a fraction of an optical cycle, providing an intrinsic few-femtosecond probe of the molecular structure and ion orbitals. In strong-field electron scattering, structural information is encoded in the momentum-transfer dependence of elastic backscattering \cite{Blaga2012Nature_LIED,Meckel2008Science_LIETD,Lin2010JPB_SelfImaging,DeGiovannini2023JPB_Perspectives,Blaga2023RSC_LIED_Chapter,Chirvi2024StructDyn_Review_FABLES}. Fixed-angle broadband laser-driven electron scattering (FABLES) in particular provides a compact route for recovering this information from a single one-dimensional rescattering spectrum at fixed backscattering angle, avoiding the polarization scans typically required in conventional laser-induced electron   diffraction (LIED)~\cite{Zuo1996CPL_LIED,Blaga2012Nature_LIED,Blaga2023RSC_LIED_Chapter,Fuest2019PRL_C60_FABLES,xu2012_cand36}.

We bring the two-color phase-control concept into strong-field diffraction in a regime where the weak $2\omega$ field is used to near-resonantly drive the ionic wave packet during the electron's continuum excursion. The key novel aspect of this work is that this field drives coupled electron--nuclear dynamics in the ion, generating phase-dependent bond-length changes that are encoded in the diffraction signal. Scanning the relative phase controls the ionic evolution, providing a phase-sensitive structural observable that, through comparison with wave-packet simulations, enables reconstruction of the initial ionic-state populations and relative phase prepared by SFI. To our knowledge, this is the first strong-field diffraction approach in which a phase-controlled auxiliary field actively drives ionic-state dynamics and the resulting structural response is used to reconstruct the ionic superposition prepared by strong-field ionization, rather than the auxiliary field serving primarily for spatial alignment~\cite{Pullen2015NatCommun_8262,Wolter2016Science_Acetylene}.

\begin{figure}[htbp]
    \centering
    \includegraphics[width=1\linewidth]{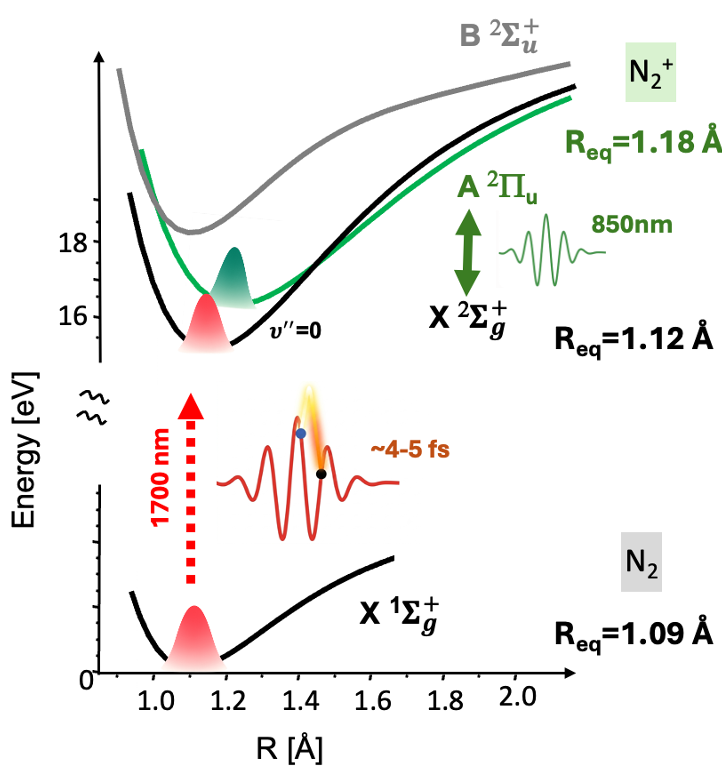}
    \caption{\textbf{Schematic of two-color phase-controlled FABLES.}
 A strong 1700-nm ($\omega$) field ionizes \ce{N2} and launches a rescattering electron, while a weak 850-nm ($2\omega$) field couples the $X\,^2\Sigma_g^+$ and $A\,^2\Pi_u$ states of \ce{N2+}. The returning electron probes the laser-driven internuclear separation, and scanning the relative phase $\phi$ provides a structural observable sensitive to the initially prepared ionic superposition.}
    \label{two-color FABLES idea}
\end{figure}

Experimentally, we use a strong 1700-nm fundamental field to ionize $\mathrm{N}_2$ and launch the rescattering electron, while its weak 850-nm second harmonic drives near-resonant coupling between the $X\,^2\Sigma_g^+$ and $A\,^2\Pi_u$ states of $\mathrm{N}_2^+$ during the electron's continuum excursion. The 850-nm field is polarized perpendicular to the fundamental, both to efficiently drive the $X$--$A$ transition in the bond-aligned ensemble selected by strong-field ionization and to minimize perturbation of the on-axis rescattering electron. This wavelength pairing therefore combines near-resonant coupling between electronic states to drive substantial dynamics in the ion on the sub-laser-cycle time scale with a high-energy rescattering plateau suitable for sub-\AA{} structural retrieval. Figure~\ref{two-color FABLES idea} illustrates the experimental concept, in which the 1700-nm field launches the rescattering electron while the weak 850-nm field drives the ionic dynamics probed at recollision.


The phase-controlled $\omega$--$2\omega$ field was generated from 60-fs, 1700-nm pulses at a 1\,kHz repetition rate and focused into a custom double time-of-flight spectrometer developed as an end station for the NeXUS facility~\cite{nexus}. The measurements reported here were performed during commissioning of the end station prior to its installation at NeXUS.  The 850-nm field was generated in a 0.5-mm Type-I BBO crystal, with the relative phase $\phi$ controlled by a birefringent calcite plate (see Sec.~S1 and Fig.~S1 of the Supplemental Material).

Here $\phi$ denotes the phase of the fundamental relative to the second harmonic (Eqs.~(S1)--(S2) of the Supplemental Material). The phase was calibrated in situ from the modulation of the high-energy rescattering cutoff by comparison with the two-color semiclassical trajectory model, providing the mapping from calcite angle to $\phi$ at the interaction region (see Supplemental Material and Figs.~S6--S7). The fundamental intensity was $I_{\omega}\simeq2.7\times10^{14}\,\mathrm{W/cm^2}$, while $I_{2\omega}/I_{\omega}=1\%$--$4\%$. At each phase step, $\mathrm{N}_2$ and Ar spectra were recorded under identical conditions, with Ar serving as an atomic reference.

At each relative phase $\phi_j$, we record energy-resolved photoelectron spectra $Y(E,\phi_j)$ for \ce{N2} and \ce{Ar} along the polarization direction of the 1700-nm fundamental field under identical experimental conditions; the corresponding raw spectra are shown in Fig.~S8 of the Supplemental Material.

\begin{figure}[htbp]
    \centering
    \includegraphics[width=1\linewidth]{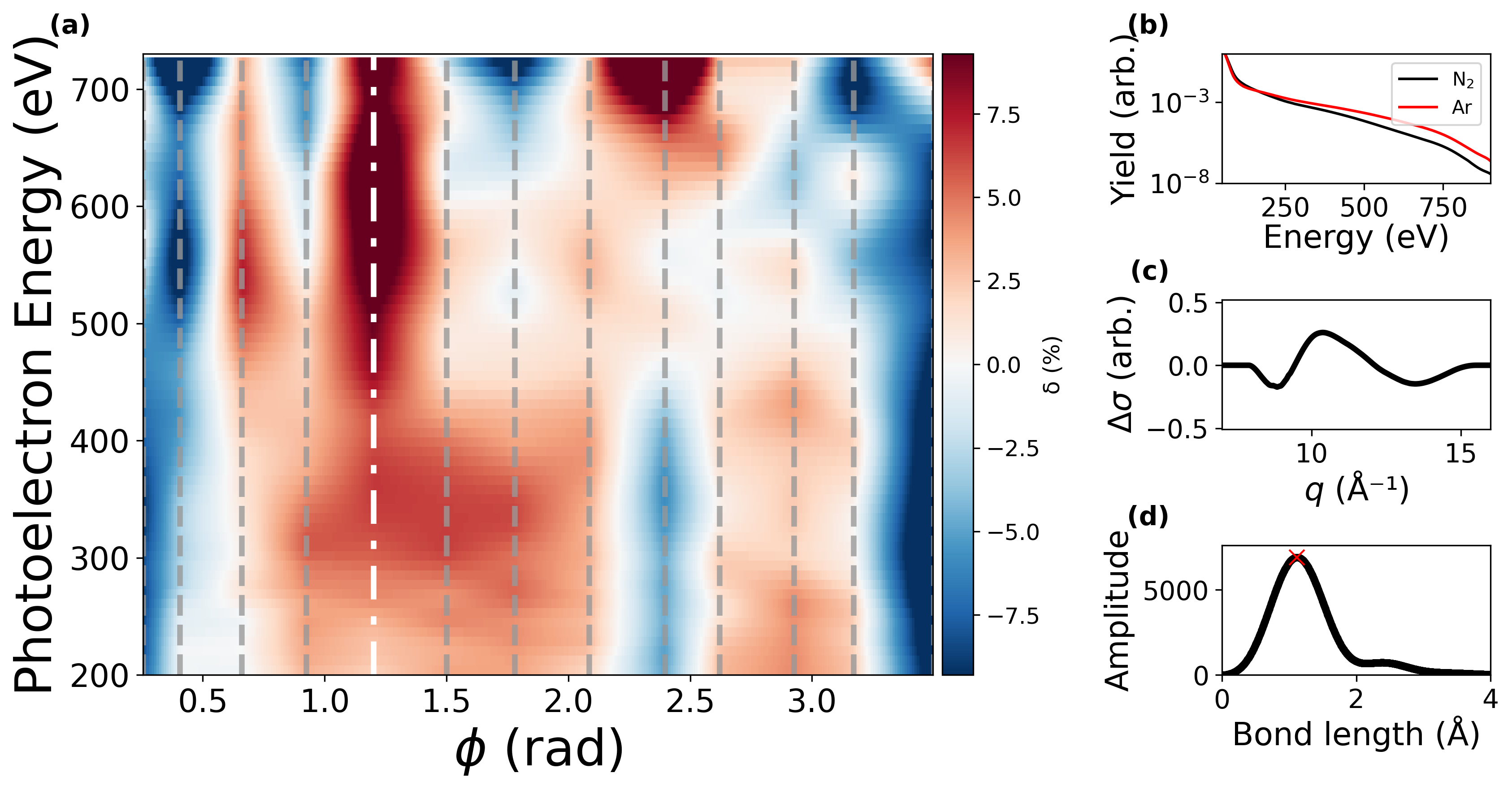}
    \caption{\textbf{Phase-dependent observable and structural retrieval.}
(a) Fractional deviation \(\delta_{\%}(E,\phi)\) of the Ar-normalized \(N_2\) photoelectron yield from its phase-averaged value. Vertical gray dashed lines indicate the sampled relative phases, and the white dashed line marks the representative phase \(\phi=1.2\) rad used in panels (b)--(d). (b) Measured \(N_2\) and Ar spectra. (c) Molecular interference signal plotted versus momentum transfer \(q\), obtained after Ar normalization, background subtraction, and semiclassical \(E\)-to-\(q\) mapping. (d) Fourier magnitude of the interference signal, whose dominant peak gives the retrieved internuclear separation.}
    \label{Analysis}
\end{figure}

Owing to its similar ionization potential, \ce{Ar} provides an atomic reference without two-center molecular interference~\cite{Xu2014NatCommun_FABLES}; normalization to the Ar spectrum also allows us to correct for small, slow experimental drifts. We define the Ar-normalized ratio as $R(E,\phi) = Y_{\mathrm{N_2}}(E,\phi)/Y_{\mathrm{Ar}}(E,\phi)$ and its phase average as $\overline{R}(E) = N_\phi^{-1}\sum_{j=1}^{N_\phi} R(E,\phi_j)$, where the sum runs over the $N_\phi$ sampled relative phases [dashed lines in Fig.~\ref{Analysis}(a)]. The phase-dependent fractional deviation is then

\begin{equation}
\delta_{\%}(E,\phi)
=
\left[
\frac{R(E,\phi)}{\overline{R}(E)}-1
\right]\times100.
\label{eq:delta}
\end{equation}

The resulting map [Fig.~\ref{Analysis}(a)] exhibits pronounced alternating enhancement and suppression across the high-energy rescattering spectrum ($200$--$1000$\,eV). Figure~\ref{Analysis}(b) shows representative full \ce{N2} and \ce{Ar} photoelectron spectra, including the low-energy direct-electron signal; only the high-energy rescattering region is used for the structural analysis. Figures~\ref{Analysis}(c) and \ref{Analysis}(d) summarize the molecular-interference and bond-length retrieval.

Structural information is encoded in the field-free elastic differential cross section (DCS), probed here by high-energy rescattered electrons. To determine the corresponding momentum transfer $q(E,\phi)$, we use a semiclassical trajectory mapping (see details in Section S2 of the Supplemental Material) in the combined $\omega$--$2\omega$ field. The semiclassical trajectory model uses a tunnel-exit radius of $R = 6.5\,\mathrm{a.u.}$, corresponding to  $I_p/E_0$ at the experimental intensity; the retrieval  is robust to this choice over the range $R = 3$--$8\,\mathrm{a.u.}$ (see Supplemental Material Fig.~S12). For each detected energy in the rescattering plateau, $3U_p \le E \le E_{\mathrm{cutoff}}(\phi)$, we identify the electron birth and return times and compute the momenta immediately before and after scattering. Thus, their difference defines the momentum transfer $q$.

To isolate the molecular diffraction signal, we analyze the data within the quantitative rescattering (QRS) framework~\cite{chen2009a_cand34}, in which the measured photoelectron yield factorizes as $Y_T(E,\phi)=\mathrm{RWP}_T(E,\phi)\sigma_T(E,\theta)$, where $\mathrm{RWP}_T$ is the returning-electron wave packet and $\sigma_T$ is the elastic differential cross section (DCS) of target $T$. Because the orthogonal $2\omega$ field introduces a small transverse momentum, the effective scattering angle $\theta\equiv\theta(E,\phi)$ deviates slightly from strict backscattering and is evaluated for each energy and phase using the semiclassical trajectory model. Using \ce{Ar} as an atomic reference~\cite{Xu2014NatCommun_FABLES}, the similar ionization potentials of \ce{N2} and Ar allow their returning-electron wave packets to approximately cancel in the measured yield ratio, $R(E,\phi)\equiv Y_{\mathrm{N_2}}(E,\phi)/Y_{\mathrm{Ar}}(E,\phi) \approx \sigma_{\mathrm{N_2}}(E,\theta)/\sigma_{\mathrm{Ar}}(E,\theta)$. The molecular DCS is therefore recovered as $\sigma_{\mathrm{N_2}}(E,\phi)\approx R(E,\phi)\sigma_{\mathrm{Ar}}(E,\theta)$. We then isolate the two-center molecular-interference term by subtracting the independent-atom contribution, $\Delta\sigma(E,\phi)\equiv\sigma_{\mathrm{N_2}}(E,\phi)-2\sigma_{\mathrm{N}}(E,\theta)$, where the atomic N and Ar differential elastic-scattering cross sections,$d\sigma/d\Omega$, are taken from the NIST Electron Elastic-Scattering Cross-Section Database ~\cite{NIST_SRD64}. Finally, mapping the detected energy onto a common momentum-transfer axis, $q=q(E,\phi)$, yields $\Delta\sigma(q,\phi)$ (Fig.~\ref{Analysis}(c)), whose Fourier transform gives the bond-length distribution (Fig.~\ref{Analysis}(d)). A summary of the analysis can also be found in Fig.~S9 of the Supplemental Material.

The magnitude of the kinematic correction was evaluated using the two-dimensional semiclassical trajectory model described in the Supplemental Material. Over the calibrated field-amplitude range used in the retrieval, $\alpha \equiv E_{2\omega}/E_{\omega}\in[0.10,0.20]$, the phase-dependent shifts of the return and scattered cutoffs remain at the few-percent level relative to the single-color limits of $3.17U_p$ and $10U_p$, respectively (see Figs.~S3 and S4 in the Supplemental Material). This small correction is included directly in the phase-dependent $E$-to-$q$ mapping. Importantly, applying a single-color $E$-to-$q$ mapping to the same two-color data shifts the absolute bond lengths but leaves the phase-dependent modulation essentially unchanged (Fig.~S13 in the Supplemental Material), demonstrating that the observed modulation is robust against the kinematic correction.

\begin{figure}[htbp]
    \centering
    \includegraphics[width=1\linewidth]{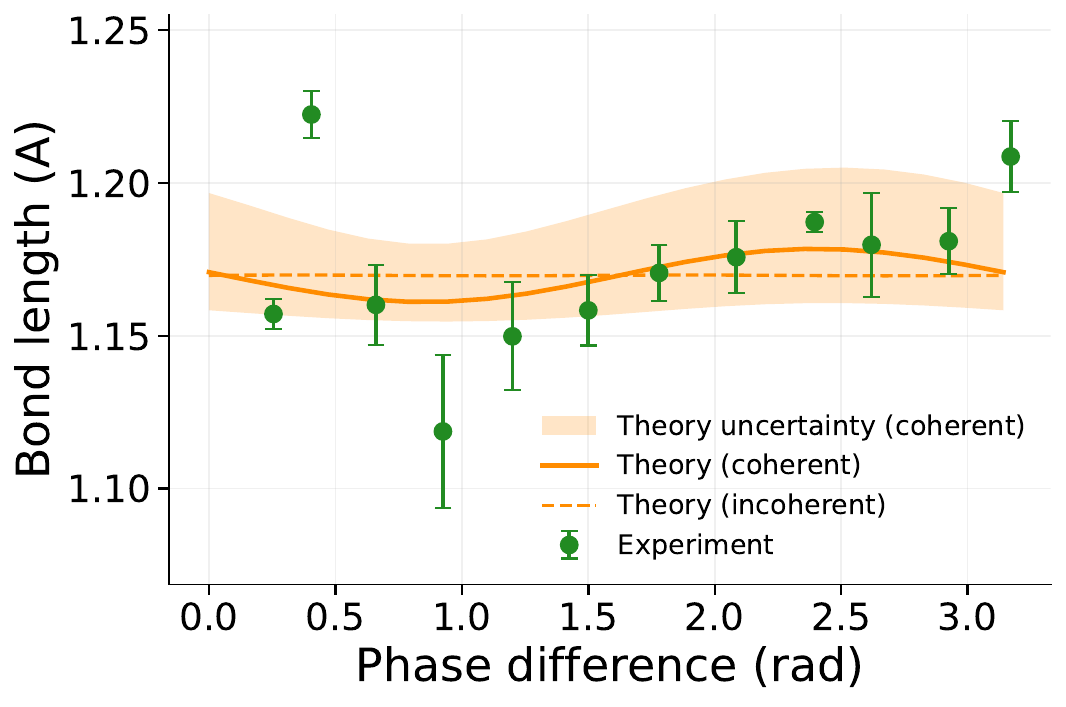}
    \caption{\textbf{Phase-dependent bond length retrieved from two-color rescattering.}
 Green points show the \ce{N2+} internuclear separation extracted from the dominant peak of $|\mathcal{F}\{\Delta\sigma(q,\phi)\}|$; error bars are one standard deviation over the retrieval uncertainty. The orange curve and shaded band show TDSE predictions for a coherently prepared $X\,^2\Sigma_g^+/A\,^2\Pi_u$ superposition, while the dashed curve shows the incoherent prediction.}
    \label{Bl vs. Phase}
\end{figure}

Figure~\ref{Bl vs. Phase} shows the phase-dependent bond length extracted from the dominant peak in the Fourier magnitude of the molecular interference signal. For reference, the NIST spectroscopic constants for \ce{N2+} equilibrium bond lengths are $R_e = 1.11642$~\AA{} and $1.1749$~\AA{} for the $X^2\Sigma_g^+$ and $A^2\Pi_u$ states, respectively~\cite{NISTN2plus}. These equilibrium distances are reference geometries rather than bounds on the retrieved structure: vertical ionization prepares a vibrational wave packet on anharmonic ionic potentials, so the rescattering-weighted bond length can transiently exceed $R_e(A)$.

For the semiclassical trajectory mapping, we used a nominal fundamental intensity of $I_{\omega}=2.7\times10^{14}$~W/cm$^2$. To account for experimental uncertainties, the full structural retrieval was repeated for four second-harmonic field-amplitude ratios spanning $E_{2\omega}/E_{\omega}=0.10$--$0.20$ and including possible residual polarization non-orthogonality (see Fig.~S5 in the Supplemental Material). At each phase, the reported bond length is the mean over these retrievals, and the error bar is their standard deviation. The phase-dependent modulation is robust to these analysis choices (see Sec.~S4 in the Supplemental Material).

To interpret the phase-dependent bond-length modulation, we performed time-dependent Schr\"odinger equation (TDSE) simulations of the nuclear wavepacket dynamics on the  X$^2\Sigma_g^+$ and A$^2\Pi_u$ states of N$_2^+$ driven by the 850 nm field.  Assuming Franck-Condon-type ionization at the peak of the strong 1700-nm field, the nuclear wave function of the vibronic ground state of N$_2$, $\chi_{\textrm{g}}$, is vertically promoted to the field-polarized X and A states of the ion, weighted by channel-specific ionization amplitudes. The two-color phase delay, $\phi$, is modeled by varying the carrier-envelope phase of the 850-nm field, corresponding to rotation of the calcite plate.

The total ion wave function, 
\begin{equation}
\label{eq:coh_ionic_wf}
\nonumber
    \Psi(\boldsymbol{r},R,t;\phi)=
    \chi_{\textrm{X}}(R,t;\phi)\psi_{\textrm{X}}(\boldsymbol{r};R) 
    + \chi_{\textrm{A}}(R,t;\phi)\psi_{\textrm{A}}(\boldsymbol{r};R),
\end{equation}
is expanded in the basis of the laser-coupled electronic states $\psi_{\textrm{X}}$ and $\psi_{\textrm{A}}$ of the ion~\cite{Langhoff1987JCP}, weighted by the vibrational wave packets $\chi_{\textrm{X}}$ and $\chi_{\textrm{A}}$, $\boldsymbol{r}$ stands for all electronic coordinates, and $R$ is the internuclear distance.
The nuclear TDSE including the laser-ion coupling (see Supplemental Material) is solved for a range of two-color phase delays and the initial conditions $\chi_{\textrm{X}}(R,t=0)=\sqrt{\mathit{w}_{\textrm{X}}}\chi_{\textrm{g}}(R)$ 
and $\chi_{\textrm{A}}(R,t=0)=\textrm{e}^{i\phi_{\textrm{XA}}}\sqrt{\mathit{w}_{\textrm{A}}}\chi_{\textrm{g}}(R)$, where $\mathit{w}_{\mathrm{X,A}}$ are the state-dependent ionization probabilities that determine the initial X to A population ratio and $\phi_{\textrm{XA}}$ determines the initial relative X-A phase created at the ionization step. Additional details of the simulations are given in the Supplemental Material.

The orange curve in Fig.~\ref{Bl vs. Phase} shows the calculated two-color phase-dependent bond length,
\begin{equation}
   \langle {R} \rangle(\phi)=\frac{ P_{\textrm{X}}(\phi)\langle R\rangle_{\textrm{X}}(\phi) + P_{\textrm{A}}(\phi)\langle R\rangle_{\textrm{A}}(\phi)}{P_{\textrm{X}}(\phi)+P_{\textrm{A}}(\phi)},
\end{equation}
where $P_{\textrm{i}}(\phi)= \int |\chi_{\textrm{i}}(R;\phi)|^2 dR$ and 
$ \langle {R} \rangle_{i}(\phi) = \int R|\chi_{\textrm{i}}(R;\phi)|^2 dR /P_{\textrm{i}}(\phi)$ (i=X,A) are the populations and bond lengths of X and A averaged over the rescattering time window of the returning electron.

Note that only in the case of initial X-A coherence, the measured bond length exhibits a distinct two-color phase-dependent modulation with $\pi$-periodicity with respect to the fundamental ($2\pi$-periodicity in the 850-nm field). The initial cross term between $\chi_{\textrm{X}}(t=0)$ and $\chi_{\textrm{A}}(t=0)$ interferes through the field coupling - the field-driven population transfer depends on the initial relative phase between the two components.
Conceptually, the experiment utilizes a homodyne-like measurement scheme.
The 850-nm field drives population transitions from the X to the A state, which interferes with the A-state population generated through direct tunneling from neutral nitrogen. Controlling the two-color phase allows us to manipulate the relative phase between these pathways and map out the initial ionic superposition.

In case of no initial X-A coherence, the system is described by a statistical mixture of the ion initially populated exclusively in (i) the X state (SFI-X) and (ii) the A state (SFI-A) using the density matrix 
\begin{eqnarray}
\nonumber
    \rho (t) = &&\mathit{w}_{\mathrm{X}}|\Psi^{(\textrm{SFI-X})}(t)\rangle \langle \Psi^{(\textrm{SFI-X})}(t)| \\
    &+& \mathit{w}_{\mathrm{A}}|\Psi^{(\textrm{SFI-A})}(t)\rangle \langle \Psi^{(\textrm{SFI-A})}(t)|.
\end{eqnarray} 
Analog to the coherent case, the ion wave functions, $\Psi^{(\textrm{SFI-i})}$, are each expanded in the basis of the laser-coupled electronic states $\psi_{\textrm{X}}$ and $\psi_{\textrm{A}}$ and weighted by the vibrational wave packets $\chi^{(\textrm{SFI-i})}_{\textrm{X}}$ and $\chi^{(\textrm{SFI-i})}_{\textrm{A}}$, but the initial conditions for the nuclear wave packet simulations are given by
$\chi^{\textrm{(SFI-X)}}_{\textrm{X}}(R,t=0)=\chi_{\textrm{g}}(R)$ 
and 
$\chi^{\textrm{(SFI-X)}}_{\textrm{A}}(R,t=0)=0$ for $\Psi^{(\textrm{SFI-X})}(t)$, and 
$\chi^{\textrm{(SFI-A)}}_{\textrm{X}}(R,t=0)=0$ 
and 
$\chi^{\textrm{(SFI-A)}}_{\textrm{A}}(R,t=0)=\chi_{\textrm{g}}(R)$ for $\Psi^{(\textrm{SFI-A})}(t)$.

The incoherent single-initial-state model fully includes the coherence the 850-nm field generates during the ion's evolution and thus allows us to isolate the effect of initial electronic coherence on the measured bond length: when starting in a \textit{single} electronic state, there is no initial coherence to beat against, the populations and ⟨R⟩ are all strictly $\phi$-independent, see Fig.~\ref{Bl vs. Phase}. For an incoherent ensemble of X and A ions, the rescattering electrons probe solely the mean bond length of the ions at the time of return. On the other hand, the coherent superposition (Eq. (\ref{eq:coh_ionic_wf})) is sensitive to the two-color phase delay, with the initial relative phase between $\chi_{\textrm{X}}$ and $\chi_{\textrm{A}}$ providing a reference against $\phi$.
 
Importantly, as $\phi_{\textrm{XA}}$ varies, the bond-length modulation exhibits an overall shift relative to the two-color phase (see Supplemental Material). The measured structural oscillation thus provides a clear fingerprint of initial electronic coherence upon SFI, and the baseline crossing of the bond-length modulation a sensitive diagnostic of the initial relative ionization phase. 
Changes in the initial population ratio and/or in the coupling field strength affect the population transfer between X and A and can thus lead to a shift in the baseline of the measured bond-length modulation toward shorter or longer bonds. They also affect the amplitude of the modulation (see Supplemental Material), but the crossing of the incoherent baseline remains unchanged.  

Within the experimental and model uncertainties, best agreement between measured and calculated $\phi$-dependent bond-length is obtained for   $P_{\textrm{X}}:P_{\textrm{A}}=5:1$ and $\phi_{\textrm{XA}}=0.75\pi$.  The shaded band around the orange curve in Fig.~\ref{Bl vs. Phase} reflects the combined uncertainty in the initial ionic state populations and in the state-dependent recollision weights. The latter is set primarily by the X- and A-channel elastic cross sections, which determine how each state contributes to the measured diffraction signal and thus also affect only the average bond length (the incoherent baseline of the modulation) and the amplitude of the modulation, but not the incoherent-coherent crossing, see Supplemental Material. The same applies to the effect of additional ionization channels, in particular the B$^2\Sigma_u^+$ state, which has been shown to play a role in SFI of N$_2$.
The reconstructed $\phi_{\textrm{XA}}$ is consistent with the results obtained via HHG spectroscopy~\cite{Mairesse2010PRL} and most likely relates to the positive (symmetric) and negative (antisymmetric) superpositions of the field-polarized (dressed) ion states, with ionization into A linked to the latter.  

\vspace{1em}

The agreement between experiment and theory supports the interpretation that the measured modulation originates from laser-driven ionic-state dynamics during the rescattering excursion, rather than from a purely kinematic two-color trajectory effect. More generally, the method can be extended to any target in which an auxiliary field modifies the ionic wave packet during the recollision interval, either through resonant population transfer or through off-resonant light-induced dressing. In such cases, the resulting electronic and nuclear response is encoded in the diffraction signal, providing a structural route to reconstructing ionic-state amplitudes and phases prepared by strong-field ionization.

In conclusion, we have introduced a two-color strong-field scattering approach in which a weak second color actively controls ionic-state dynamics during the electron's continuum excursion, while the rescattering electron provides a structural readout of the resulting response. By scanning the relative phase and using an atomic reference to isolate the two-center interference, we retrieve phase-dependent sub-\AA{} bond-length changes in \ce{N2+}. Combined with TDSE simulations, these measurements enable reconstruction of the amplitudes and relative phase of the ionic superposition prepared by strong-field ionization. Unlike earlier HHG-based reconstruction approaches~\cite{Smirnova2009Nature,Mairesse2010PRL,Bruner2016FD}, this method does not require energy- and channel-resolved phase-coherent recombination amplitudes and phases.

Looking forward, our approach could be extended to heteronuclear molecular ions such as CO$^+$, where the weak field overlaps the $A\,^2\Pi-X\,^2\Sigma^+$ comet-tail system and could coherently drive or dress the ionic wave packet during rescattering~\cite{Zhang2014_COplus}. More generally, tuning the second color on or off resonance could distinguish real population transfer from virtual light-induced dressing, including transient Stark shifts and modifications of the ionic potential-energy landscape~\cite{Sussman2006,Sindelka2011}. For larger molecules, an auxiliary alignment pulse could define the molecular frame, enabling molecular-frame FABLES measurements of selected bond projections, while extension to two-color laser-induced electron diffraction (LIED) could provide the full momentum-resolved diffraction map. In these extensions, the second color would enable active reconstruction of the ionic-state amplitudes and relative phases prepared by strong-field ionization, providing quantitative benchmarks for theories of non-equilibrium molecular-ion dynamics.

This work was supported by the National Science Foundation (NSF grant No.\,1935885) and the U.S. Department of Energy, Office of Science, Basic Energy Sciences, under Award No. DE-FG02-04ER15614. M.R. and F.M. gratefully acknowledge stimulating discussions with Serguei Patchkovskii. C.I.B. acknowledges support from Chemical Sciences, Geosciences and Biosciences Division, Office of Basic Energy Sciences, Office of Science, U.S. Department of Energy Grant No. DE-FG02-86ER13491.

\bibliographystyle{apsrev4-1}  
\bibliography{bib}

@article{Porat2018NatComm,
  author  = {Porat, G. and Alon, G. and Rozen, S. and Pedatzur, O. and Krüger, M. and Azoury, D. and Natan, A. and Orenstein, G. and Bruner, B. D. and Vrakking, M. J. J. and Dudovich, N.},
  title   = {Attosecond time-resolved photoelectron holography},
  journal = {Nature Communications},
  year    = {2018},
  volume  = {9},
  pages   = {2805},
  doi     = {10.1038/s41467-018-05185-6},
  url     = {https://doi.org/10.1038/s41467-018-05185-6}
}

@article{Mairesse2010PRL,
  title = {High Harmonic Spectroscopy of Multichannel Dynamics in Strong-Field Ionization},
  author = {Mairesse, Y. and Higuet, J. and Dudovich, N. and Shafir, D. and Fabre, B. and M\'evel, E. and Constant, E. and Patchkovskii, S. and Walters, Z. and Ivanov, M. Yu. and Smirnova, O.},
  journal = {Phys. Rev. Lett.},
  volume = {104},
  issue = {21},
  pages = {213601},
  numpages = {4},
  year = {2010},
  month = {May},
  publisher = {American Physical Society},
  doi = {10.1103/PhysRevLett.104.213601},
  url = {https://link.aps.org/doi/10.1103/PhysRevLett.104.213601}
}

@article{Chaloupka2016PRL,
  title = {Dynamics of Strong-Field Double Ionization in Two-Color Counterrotating Fields},
  author = {Chaloupka, Jan L. and Hickstein, Daniel D.},
  journal = {Phys. Rev. Lett.},
  volume = {116},
  issue = {14},
  pages = {143005},
  numpages = {5},
  year = {2016},
  month = {Apr},
  publisher = {American Physical Society},
  doi = {10.1103/PhysRevLett.116.143005},
  url = {https://link.aps.org/doi/10.1103/PhysRevLett.116.143005}
}

@Article{xu2012_cand36,
  author    = {Xu, Junliang and Blaga, Cosmin I. and DiChiara, Anthony D. and Sistrunk, Emily and Zhang, Kaikai and Chen, Zhangjin and Le, Anh-Thu and Morishita, Toru and Lin, C. D. and Agostini, Pierre and DiMauro, Louis F.},
  journal   = {Physical Review Letters},
  title     = {Laser-{Induced} {Electron} {Diffraction} for {Probing} {Rare} {Gas} {Atoms}},
  year      = {2012},
  month     = dec,
  number    = {23},
  pages     = {233002},
  volume    = {109},
  doi       = {10.1103/PhysRevLett.109.233002},
  publisher = {American Physical Society},
  urldate   = {2025-03-05},
}

@article{Jin2024UltrafastScience,
author = {Wuwei Jin  and Tao Jiang  and Jinlei Liu  and Sizuo Luo  and Dianxiang Ren  and Xiaokai Li  and Chuncheng Wang  and Yue Lang  and Xiaowei Wang  and Jing Zhao  and Zengxiu Zhao  and Dajun Ding },
title = {Strong Field Ionization Dynamics Resolved by Two-Color Elliptical Phase-of-Phase Spectroscopy},
journal = {Ultrafast Science},
volume = {4},
number = {},
pages = {0066},
year = {2024},
doi = {10.34133/ultrafastscience.0066},
URL = {https://spj.science.org/doi/abs/10.34133/ultrafastscience.0066},
eprint = {https://spj.science.org/doi/pdf/10.34133/ultrafastscience.0066}
}

@misc{nexus,
  author       = {{NeXUS Facility}},
  title        = {NeXUS: National eXtreme Ultrafast Science Facility},
  howpublished = {\url{https://nsf-nexus.osu.edu/}},
  note         = {Accessed: 16 Jun 2025},
  year         = {2025},
}

@Article{chen2009a_cand34,
  author    = {Chen, Zhangjin and Le, Anh-Thu and Morishita, Toru and Lin, C. D.},
  journal   = {Physical Review A},
  title     = {Quantitative {Rescattering} {Theory} for {Laser-Induced} {High-Energy} {Plateau} {Photoelectron} {Spectra}},
  year      = {2009},
  month     = mar,
  number    = {3},
  pages     = {033409},
  volume    = {79},
  doi       = {10.1103/PhysRevA.79.033409},
  publisher = {American Physical Society},
  urldate   = {2025-03-05},
}

@article{Zuo1996CPL_LIED,
  author  = {Zuo, T. and Bandrauk, A. D. and Corkum, P. B.},
  title   = {Laser-induced electron diffraction: A new tool for probing ultrafast molecular dynamics},
  journal = {Chemical Physics Letters},
  year    = {1996},
  volume  = {259},
  number  = {3-4},
  pages   = {313--320},
  doi     = {10.1016/0009-2614(96)00786-5},
}

@article{Meckel2008Science_LIETD,
  author  = {Meckel, M. and Kieffer, J. C. and D{\"o}rner, R. and Villeneuve, D. M. and Comtois, D. and Zeidler, D. and Staudte, A. and Pavi{\v{c}}i{\'c}, D. and Bandulet, H. C. and P{\'e}pin, H. and Corkum, P. B.},
  title   = {Laser-Induced Electron Tunneling and Diffraction},
  journal = {Science},
  year    = {2008},
  volume  = {320},
  number  = {5882},
  pages   = {1478--1482},
  doi     = {10.1126/science.1157980},
}

@article{Lin2010JPB_SelfImaging,
  author  = {Lin, C. D. and Le, A.-T. and Chen, Z. and Morishita, T. and Lucchese, R.},
  title   = {Strong-field rescattering physics---Self-imaging of a molecule by its own electrons},
  journal = {Journal of Physics B: Atomic, Molecular and Optical Physics},
  year    = {2010},
  volume  = {43},
  number  = {12},
  pages   = {122001},
  doi     = {10.1088/0953-4075/43/12/122001},
}

@article{Blaga2012Nature_LIED,
  author  = {Blaga, C. I. and Xu, J. and DiChiara, A. D. and Sistrunk, E. and Zhang, K. and Agostini, P. and Miller, T. A. and DiMauro, L. F. and Lin, C. D.},
  title   = {Imaging ultrafast molecular dynamics with laser-induced electron diffraction},
  journal = {Nature},
  year    = {2012},
  volume  = {483},
  number  = {7388},
  pages   = {194--197},
  doi     = {10.1038/nature10820},
}

@article{Pullen2015NatCommun_8262,
  author  = {Pullen, Michael G. and Wolter, Benjamin and Le, Anh-Thu and Baudisch, Matthias and Hemmer, Micha{\"e}l and Senftleben, Arne and Schr{\"o}ter, Claus Dieter and Ullrich, Joachim and Moshammer, Robert and Lin, C. D. and Biegert, Jens},
  title   = {Imaging an aligned polyatomic molecule with laser-induced electron diffraction},
  journal = {Nature Communications},
  year    = {2015},
  volume  = {6},
  pages   = {7262},
  doi     = {10.1038/ncomms8262},
}

@article{Wolter2016Science_Acetylene,
  author  = {Wolter, B. and Pullen, M. G. and Le, A.-T. and Baudisch, M. and Doblhoff-Dier, K. and Senftleben, A. and Hemmer, M. and Schr{\"o}ter, C. D. and Ullrich, J. and Pfeifer, T. and Moshammer, R. and Gr{\"a}fe, S. and Vendrell, O. and Lin, C. D. and Biegert, J.},
  title   = {Ultrafast electron diffraction imaging of bond breaking in di-ionized acetylene},
  journal = {Science},
  year    = {2016},
  volume  = {354},
  number  = {6310},
  pages   = {308--312},
  doi     = {10.1126/science.aah3429},
}

@article{DeGiovannini2023JPB_Perspectives,
  author  = {De Giovannini, Umberto and K{\"u}pper, Jochen and Trabattoni, Andrea},
  title   = {New perspectives in time-resolved laser-induced electron diffraction},
  journal = {Journal of Physics B: Atomic, Molecular and Optical Physics},
  year    = {2023},
  volume  = {56},
  pages   = {054002},
  doi     = {10.1088/1361-6455/acb872},
}

@incollection{Blaga2023RSC_LIED_Chapter,
  author    = {Blaga, Cosmin I.},
  title     = {Laser Induced Electron Diffraction},
  booktitle = {Structural Dynamics with X-ray and Electron Scattering},
  editor    = {Amini, K. and Rouz{\'e}e, A. and Vrakking, M. J. J.},
  publisher = {Royal Society of Chemistry},
  year      = {2023},
  volume    = {25},
  chapter   = {13},
  pages     = {511--534},
}

@article{Xu2014NatCommun_FABLES,
  author  = {Xu, Junliang and Blaga, Cosmin I. and Zhang, Kaikai and Lai, Yu Hang and Lin, C. D. and Miller, Terry A. and Agostini, Pierre and DiMauro, Louis F.},
  title   = {Diffraction using laser-driven broadband electron wave packets},
  journal = {Nature Communications},
  year    = {2014},
  volume  = {5},
  pages   = {4635},
  doi     = {10.1038/ncomms5635},
}

@article{Fuest2019PRL_C60_FABLES,
  author  = {Fuest, Harald and Lai, Yu Hang and Blaga, Cosmin I. and Suzuki, Kazuma and Xu, Junliang and Rupp, Philipp and Li, Hui and Wnuk, Pawel and Agostini, Pierre and Yamazaki, Kaoru and Kanno, Manabu and Kono, Hirohiko and Kling, Matthias F. and DiMauro, Louis F.},
  title   = {Diffractive Imaging of C$_{60}$ Structural Deformations Induced by Intense Femtosecond Midinfrared Laser Fields},
  journal = {Physical Review Letters},
  year    = {2019},
  volume  = {122},
  number  = {5},
  pages   = {053002},
  doi     = {10.1103/PhysRevLett.122.053002},
}

@article{Chirvi2024StructDyn_Review_FABLES,
  author  = {Chirvi, K. and Biegert, J.},
  title   = {Laser-induced electron diffraction: Imaging of a single gas-phase molecular structure with one of its own electrons},
  journal = {Structural Dynamics},
  year    = {2024},
  volume  = {11},
  number  = {4},
  pages   = {041301},
  doi     = {10.1063/4.0000237},
}

@article{Brugnera2011PRL_TwoColorTrajectories,
  author  = {Brugnera, Leonardo and Hoffmann, David J. and Siegel, Thomas and Frank, Felix and Za{\"i}r, Amelle and Tisch, John W. G. and Marangos, Jonathan P.},
  title   = {Trajectory Selection in High Harmonic Generation by Controlling the Phase between Orthogonal Two-Color Fields},
  journal = {Physical Review Letters},
  year    = {2011},
  volume  = {107},
  pages   = {153902},
  doi     = {10.1103/PhysRevLett.107.153902}
}

@article{Niikura2010PRL_TwoColorOrbitalSymmetry,
  author  = {Niikura, Hiromichi and Dudovich, Nirit and Villeneuve, D. M. and Corkum, P. B.},
  title   = {Mapping Molecular Orbital Symmetry on High-Order Harmonic Generation Spectrum Using Two-Color Laser Fields},
  journal = {Physical Review Letters},
  year    = {2010},
  volume  = {105},
  pages   = {053003},
  doi     = {10.1103/PhysRevLett.105.053003}
}

@article{Zipp2014,
  author = {Zipp, Lucas J. and Natan, Adi and Bucksbaum, Philip H.},
  title = {Probing electron delays in above-threshold ionization},
  journal = {Optica},
  volume = {1},
  number = {6},
  pages = {361--364},
  year = {2014},
  doi = {10.1364/OPTICA.1.000361}
}

@article{Skruszewicz2015,
  author = {Skruszewicz, S. and Tiggesb{\"a}umker, J. and Meiwes-Broer, K.-H. and Arbeiter, M. and Fennel, Th. and Bauer, D.},
  title = {Two-Color Strong-Field Photoelectron Spectroscopy and the Phase of the Phase},
  journal = {Physical Review Letters},
  volume = {115},
  pages = {043001},
  year = {2015},
  doi = {10.1103/PhysRevLett.115.043001}
}

@article{Smirnova2009Nature,
  author  = {Smirnova, Olga and Mairesse, Yann and Patchkovskii, Serguei and Dudovich, Nirit and Villeneuve, David M. and Corkum, Paul B. and Ivanov, Misha Yu.},
  title   = {High harmonic interferometry of multi-electron dynamics in molecules},
  journal = {Nature},
  volume  = {460},
  number  = {7258},
  pages   = {972--977},
  year    = {2009},
  doi     = {10.1038/nature08253}
}

@article{Bruner2016FD,
  author  = {Bruner, Barry D. and Ma{\v{s}}{\'i}n, Zden{\v{e}}k and Negro, Matteo and Morales, Felipe and Brambila, Danilo and Devetta, Michele and Faccial{\`a}, Davide and Harvey, Alex G. and Ivanov, Misha and Mairesse, Yann and Patchkovskii, Serguei and Serbinenko, Valeria and Soifer, Hadas and Stagira, Salvatore and Vozzi, Caterina and Dudovich, Nirit and Smirnova, Olga},
  title   = {Multidimensional high harmonic spectroscopy of polyatomic molecules: detecting sub-cycle laser-driven hole dynamics upon ionization in strong mid-{IR} laser fields},
  journal = {Faraday Discussions},
  volume  = {194},
  pages   = {369--405},
  year    = {2016},
  doi     = {10.1039/C6FD00130K}
}

@article{McFarland2008Science,
  author  = {McFarland, Brian K. and Farrell, Joseph P. and Bucksbaum, Philip H. and G{\"u}hr, Markus},
  title   = {High harmonic generation from multiple orbitals in {N$_2$}},
  journal = {Science},
  volume  = {322},
  number  = {5905},
  pages   = {1232--1235},
  year    = {2008},
  doi     = {10.1126/science.1162780}
}

@article{Haessler2010NatPhys,
  author  = {Haessler, S. and Caillat, J. and Boutu, W. and Giovanetti-Teixeira, C. and Ruchon, T. and Auguste, T. and Diveki, Z. and Breger, P. and Maquet, A. and Carr{\'e}, B. and Ta{\"i}eb, R. and Sali{\`e}res, P.},
  title   = {Attosecond imaging of molecular electronic wavepackets},
  journal = {Nature Physics},
  volume  = {6},
  number  = {3},
  pages   = {200--206},
  year    = {2010},
  doi     = {10.1038/nphys1511}
}

@article{Peschel2022NatCommun,
  author  = {Peschel, Jasper and Busto, David and Plach, Marius and Bertolino, Mattias and Hoflund, Maria and Maclot, Sylvain and Vinbladh, Jimmy and Wikmark, Hampus and Zapata, Felipe and Lindroth, Eva and Gisselbrecht, Mathieu and Dahlstr{\"o}m, Jan Marcus and L'Huillier, Anne and Eng-Johnsson, Per},
  title   = {Attosecond dynamics of multi-channel single photon ionization},
  journal = {Nature Communications},
  volume  = {13},
  pages   = {5205},
  year    = {2022},
  doi     = {10.1038/s41467-022-32780-5}
}

@article{Liu2015PRL,
  author  = {Liu, Zuoye and Cavaletto, Stefano M. and Ott, Christian and Meyer, Kristina and Mi, Yonghao and Harman, Zolt{\'a}n and Keitel, Christoph H. and Pfeifer, Thomas},
  title   = {Phase Reconstruction of Strong-Field Excited Systems by Transient-Absorption Spectroscopy},
  journal = {Physical Review Letters},
  volume  = {115},
  pages   = {033003},
  year    = {2015},
  doi     = {10.1103/PhysRevLett.115.033003}
}

@article{Langhoff1987JCP,
  author  = {Langhoff, Stephen R. and Bauschlicher, Jr., Charles W. and Partridge, Harry},
  title   = {Theoretical study of the {N}$_2^+$ {Meinel} system},
  journal = {The Journal of Chemical Physics},
  volume  = {87},
  number  = {8},
  pages   = {4716--4721},
  year    = {1987},
  doi     = {10.1063/1.452835}
}

@article{Piper2025PRL_QTS,
  author = {Piper, Andrew J. and Liu, Qiaoyi and Camacho Garibay, Abraham and Kiesewetter, Dietrich and Leshchenko, Vyacheslav and B{\ae}kh{\o}j, Jens E. and Agostini, Pierre and Schafer, Kenneth J. and DiMauro, Louis F. and Tang, Yaguo},
  title = {Attosecond Clocking and Control of Strong Field Quantum Trajectories},
  journal = {Physical Review Letters},
  volume = {134},
  pages = {073201},
  year = {2025},
  doi = {10.1103/PhysRevLett.134.073201}
}

@article{Zhang2014_COplus,
  title={Role of the $A\,^2\Pi$ state in the strong-field ionization of CO molecules},
  author={Zhang, B. and Yuan, J. and Zhao, Z.},
  journal={Physical Review A},
  volume={90},
  number={2},
  pages={023402},
  year={2014},
  publisher={APS}
}

@article{Sussman2006,
  title={Dynamic Stark control of molecular pathways in electronic processes},
  author={Sussman, Benjamin J and Townsend, D and Ivanov, M Yu and Stolow, Albert},
  journal={Science},
  volume={314},
  number={5797},
  pages={278--281},
  year={2006},
  publisher={American Association for the Advancement of Science}
}

@article{Sindelka2011,
  title={Strong impact of light-induced conical intersections on the spectrum of diatomic molecules},
  author={{\v{S}}indelka, Milan and Moiseyev, Nimrod and Cederbaum, Lorenz S},
  journal={Journal of Physics B: Atomic, Molecular and Optical Physics},
  volume={44},
  number={4},
  pages={045603},
  year={2011},
  publisher={IOP Publishing}
}

@article{Richter2020Optica,
  author  = {Richter, Maria and Lytova, Marianna and Morales, Felipe
             and Haessler, Stefan and Smirnova, Olga and Spanner, Michael
             and Ivanov, Misha},
  title   = {Rotational Quantum Beat Lasing Without Inversion},
  journal = {Optica},
  volume  = {7},
  number  = {6},
  pages   = {586--592},
  year    = {2020},
  doi     = {10.1364/OPTICA.390665},
  url     = {https://doi.org/10.1364/OPTICA.390665}
}

@article{Yao2011,
  title = {High-brightness switchable multiwavelength remote laser in air},
  author = {Yao, Jinping and Zeng, Bin and Xu, Huailiang and Li, Guihua and Chu, Wei and Ni, Jielei and Zhang, Haisu and Chin, See Leang and Cheng, Ya and Xu, Zhizhan},
  journal = {Phys. Rev. A},
  volume = {84},
  issue = {5},
  pages = {051802(R)},
  numpages = {5},
  year = {2011},
  month = {Nov},
  publisher = {American Physical Society},
  doi = {10.1103/PhysRevA.84.051802},
  url = {https://link.aps.org/doi/10.1103/PhysRevA.84.051802}
}

@misc{NISTN2plus,
  author       = {Huber, Klaus P. and Herzberg, Gerhard H.},
  title        = {Constants of Diatomic Molecules},
  booktitle    = {NIST Chemistry WebBook, NIST Standard Reference Database Number 69},
  editor       = {Linstrom, Peter J. and Mallard, W. Gary},
  publisher    = {National Institute of Standards and Technology},
  address      = {Gaithersburg, MD},
  doi          = {10.18434/T4D303},
  url          = {https://webbook.nist.gov/cgi/cbook.cgi?ID=C13966046&Mask=3FF0},
  note         = {Nitrogen cation, retrieved August 24, 2026}
}

@article{Pabst2016PRA,
  author    = {Pabst, Stefan and Lein, Manfred and W{\"o}rner, Hans Jakob},
  title     = {Preparing Attosecond Coherences by Strong-Field Ionization},
  journal   = {Physical Review A},
  volume    = {93},
  number    = {2},
  pages     = {023412},
  year      = {2016},
  month     = feb,
  publisher = {American Physical Society},
  doi       = {10.1103/PhysRevA.93.023412},
  url       = {https://doi.org/10.1103/PhysRevA.93.023412}
}

@article{Kraus2015Science,
  author  = {Kraus, P. M. and Mignolet, B. and Baykusheva, D. and
             Rupenyan, A. and Horn{\'y}, L. and Penka, E. F. and
             Grassi, G. and Tolstikhin, Oleg I. and Schneider, J. and
             Jensen, Frank and Madsen, Lars Bojer and Bandrauk, A. D. and
             Remacle, F. and W{\"o}rner, H. J.},
  title   = {Measurement and Laser Control of Attosecond Charge Migration
             in Ionized Iodoacetylene},
  journal = {Science},
  volume  = {350},
  number  = {6262},
  pages   = {790--795},
  year    = {2015},
  month   = nov,
  doi     = {10.1126/science.aab2160},
  url     = {https://doi.org/10.1126/science.aab2160}
}

@article{Vacher2017PRL,
  author    = {Vacher, Morgane and Bearpark, Michael J. and
               Robb, Michael A. and Malhado, Jo{\~a}o Pedro},
  title     = {Electron Dynamics upon Ionization of Polyatomic Molecules:
               Coupling to Quantum Nuclear Motion and Decoherence},
  journal   = {Physical Review Letters},
  volume    = {118},
  number    = {8},
  pages     = {083001},
  year      = {2017},
  month     = feb,
  publisher = {American Physical Society},
  doi       = {10.1103/PhysRevLett.118.083001},
  url       = {https://doi.org/10.1103/PhysRevLett.118.083001}
}

@article{Zhao2017PRA,
  author    = {Zhao, Arthur and S{\'a}ndor, P{\'e}ter and
               Tagliamonti, Vincent and Rozgonyi, Tam{\'a}s and
               Marquetand, Philipp and Weinacht, Thomas},
  title     = {Ionic Dynamics Underlying Strong-Field Dissociative
               Molecular Ionization},
  journal   = {Physical Review A},
  volume    = {96},
  number    = {2},
  pages     = {023404},
  year      = {2017},
  month     = aug,
  publisher = {American Physical Society},
  doi       = {10.1103/PhysRevA.96.023404},
  url       = {https://doi.org/10.1103/PhysRevA.96.023404}
}

@misc{NIST_SRD64,
  author       = {A. Jablonski and F. Salvat and C. J. Powell and A. Y. Lee},
  title        = {NIST Electron Elastic-Scattering Cross-Section Database, Version 5.0},
  publisher    = {National Institute of Standards and Technology},
  address      = {Gaithersburg, MD},
  year         = {2023},
  note         = {Standard Reference Database 64, accessed August 28, 2026},
  url          = {https://srdata.nist.gov/srd64/}
}

\subsection*{Two-color phase calibration}

The relative phase $\phi$ between the 1700-nm fundamental and 850-nm second-harmonic fields was calibrated independently of the structural retrieval using the phase-dependent modulation of the high-energy rescattering cutoff. A high-resolution calcite-delay scan was recorded prior to the high-statistics FABLES measurements, and the measured cutoff modulation was compared with the maximum scattered-electron energy
predicted by the two-color semiclassical trajectory model. Matching the periodicity and extrema of the measured and calculated modulations provides the mapping from calcite angle to the relative phase $\phi$ at the interaction region; constant optical phase offsets are thereby absorbed into the experimental phase calibration. The calibrated phase values were subsequently used for the bond-length measurements, so the phase axis of the structural modulation is determined independently of the ionic-wave-packet reconstruction. Further details are provided in Sec.~S2.1 of the Supplemental Material.

\subsection*{Two-color kinematic correction}

A key question is whether the observed phase-dependent bond-length modulation could arise from the two-color $E$-to-$q$ mapping itself rather than from the measured molecular response. To test this possibility, we repeated the full structural retrieval with the second-harmonic amplitude set to $\alpha=0$ in the semiclassical trajectory model, corresponding to a single-color $E$-to-$q$ mapping applied to the same two-color experimental spectra. Because the spectra were acquired with both colors present, this procedure removes only the phase-dependent kinematic correction; any response induced by the 850-nm field remains encoded in the measured yields.

\begin{figure}[htbp]
    \centering
    \includegraphics[width=1\linewidth]{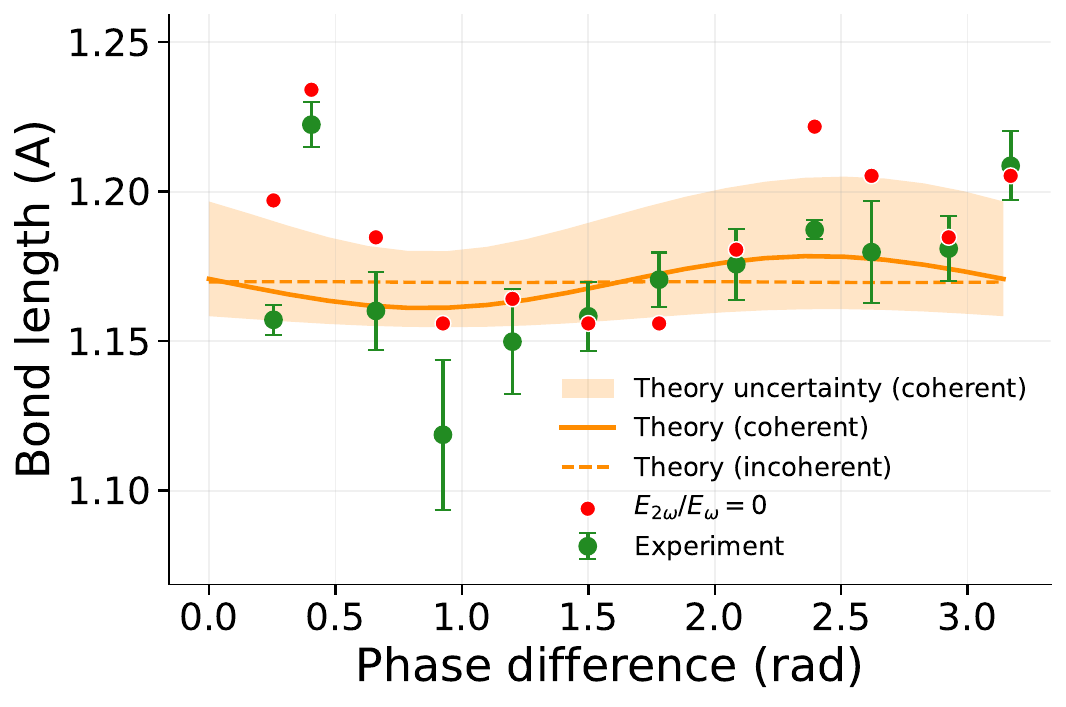}
    \caption{\textbf{Robustness against the two-color kinematic correction.}
  Green points show the nominal experimental retrieval averaged over the calibrated second-harmonic field-amplitude range and polarization mappings; red points show the retrieval obtained by applying a single-color
  ($\alpha=0$) $E$-to-$q$ mapping to the same two-color spectra.  Removing the kinematic correction shifts the absolute bond lengths but preserves the phase-dependent modulation. The solid and dashed orange curves show the TDSE predictions for coherent and incoherent initial ionic
  preparations, respectively.}
  \label{fig:kinematic_test}
\end{figure}

Figure~\ref{fig:kinematic_test} compares the nominal two-color retrieval with the result obtained using the single-color mapping. Removing the kinematic correction shifts the absolute retrieved bond lengths, as expected, but leaves the phase-dependent modulation essentially unchanged. The two analyses therefore exhibit the same modulation phase and overall shape within the experimental uncertainty. Because they are performed on the \emph{same} measured spectra, this comparison provides a data-driven demonstration that the modulation is not generated by the phase-dependent $E$-to-$q$ mapping. The TDSE predictions are shown for comparison: the phase dependence retained in both retrievals is captured by the coherent $X\,^2\Sigma_g^+/A\,^2\Pi_u$ calculation, whereas the incoherent calculation is essentially phase independent. Thus, the two-color trajectory correction refines the absolute structural retrieval but is not responsible for the observed phase-dependent modulation.

\end{document}